\documentclass[a4paper,11pt]{article}

\usepackage{pos}
\usepackage{slashed}
\usepackage{ascmac}
\usepackage{tcolorbox}
\usepackage{tikz-feynman}
    \tikzfeynmanset{warn luatex=false}
\usepackage{tikz}
\usepackage{bm}
\usepackage[capitalize]{cleveref}

\newcommand{\doublecirc}{{\ooalign{$\bigcirc$\crcr\hss$\circ$\hss}}}

\title{Quantum Wishlist: Lessons from Parton Showers}

\author*[a,b,c,d]{Masahito Yamazaki}

\affiliation[a]{Department of Physics, University of Tokyo, Tokyo 113-0033, Japan}
\affiliation[b]{Kavli Institute for Physics and Mathematics of the Universe (Kavli IPMU), \\ UTIAS, University of Tokyo, Chiba 277-8583, Japan}
\affiliation[c]{Trans-Scale Quantum Science Institute,  University of Tokyo, Tokyo 113-0033, Japan}
\affiliation[d]{Center for Data-Driven Discovery (CD3), Kavli IPMU, University of Tokyo, Chiba 277-8583, Japan}

\emailAdd{masahito.yamazaki@ipmu.jp}

\abstract{
We discuss general criteria that could guide us in applying quantum algorithms/computers to problems in high-energy physics. We then discuss the particular example of parton showers with quantum interference. We summarize the basic ideas behind the classical and quantum parton shower Monte Carlo algorithms and highlight the importance of quantum/classical hybrid algorithms. Our finding in quantum parton showers could serve as a useful case study for further exploration of quantum algorithms/computers in high-energy physics.
}

\FullConference{The XVIth Quark Confinement and the Hadron Spectrum Conference (QCHSC24)\\
 19-24 August, 2024\\
 Cairns Convention Centre, Cairns, Queensland, Australia\\}

\tableofcontents

\begin{document}
\maketitle

\section{Quantum Era in High Energy Physics}

In the history of physics, breakthroughs have often been made 
thanks to the advent of novel technologies. Due to the rapid progress of quantum technologies,
it could be perfect timing now to explore future applications of quantum algorithms and quantum computers\footnote{In this paper we do not discuss applications of quantum sensing in HEP.} to the problems in high-energy physics (HEP) \cite{Bauer:2022hpo,DiMeglio:2023nsa}---one may well argue that we are in the ``quantum era in HEP.''

When we discuss ``quantum in HEP,''
one should be aware of the two sides of the coin.
On the one hand, there are known quantum algorithms with guaranteed speedups over their classical counterparts \cite{Shor:1994jg,Grover:1996rk}. Moreover,
quantum devices today have $\gtrsim O(100)$ qubits with comparable depth, and in principle can represent more states than can be stored
even in the most advanced classical supercomputers today, and hence we are already in the ``utility era'' \cite{Kim:2023bwr}.

On the other hand, one should keep in mind that 
the qubits today are noisy \cite{Preskill:2018jim}, even after applying error mitigation techniques. While we eventually may attain practical quantum error corrections
in the future, the number of logical qubits
protected by the error corrections will still be limited
due to the huge overhead in the error correction;
it will likely be unrealistic, even say thirty years from now, e.g.\ to directly (i.e.\ in brute-force) simulate the whole scattering process of Standard Model (SM) particles at the Large Hadron Collider (LHC) in quantum computers, with full error-corrected logical qubits.

Of course, such a discussion does not prevent us from being \emph{cautiously optimistic} about the future of this fascinating topic. It does, however, seem to be necessary to choose appropriate problems 
if we realistically want to have 
some impactful physics results in the near future.

\section{Quantum Wishlist}

In view of the discussion above, I propose some possible guidelines in search of promising research directions.
As we will discuss, the guidelines are far from perfect and are somewhat biased due to my upbringing as a physicist. I hope, however,
that the general idea will be useful to many researchers in this area.

\paragraph{I. Availability of Simplified Models}

While the ultimate problem we are interested in 
can be very complicated and sophisticated,
a good strategy in physics is to start with a much-simplified model, which is simple enough but captures some essence of the physics. Such a simplified model could be tractable already in the present-day quantum devices.
We can then add layers of complications to make the discussion more realistic so that subsequent progress can be split into several manageable steps.

\paragraph{II. Scalability}
While it is important to start from simplified problems as mentioned above, the problem needs to be scalable when extrapolated to the future. In other words,
we need an estimate on how to approach a full-scale problem of physical interest
as the quantum devices improve, in the numbers of qubits and gates, the fidelities of the qubits,
as well as the expected overhead in quantum error corrections, for example.
While this is inevitably affected by the uncertainties in the future development of quantum computers,
an approximate estimate is often useful already.

\paragraph{III. Importance of Quantum Effects}
We require that
the physics problem at hand in itself involves important quantum effects,
originating e.g.\ from quantum interference and/or quantum entanglement. While it is very difficult in general to have the theoretical guarantee that the problem could be solved more efficiently by quantum devices than their classical counterparts, it seems natural to 
apply quantum computers to inherently quantum-mechanical problems.\footnote{When the problem is classical, one often needs a significant overhead for simply loading the classical data into quantum bits. Such an overhead sometimes does not exist when the physics problem in itself is quantum.} This follows the often-quoted philosophy by R.~Feynman \cite{Feynman:1981tf}.

\paragraph{IV. Interface to Classical Computations}

While quantum computers will surely improve in the future, we do not expect 
classical computers to be superseded completely by quantum computers.\footnote{At least while the author is alive.} There are often many fantastic existing classical algorithms
already fine-tuned to the problem, in which case we should not try to throw them away and start from scratch. Rather, we need to come up with an interface between existing classical algorithms to the newly-developed quantum algorithms, and the 
eventual algorithm will inevitably be a hybrid of
quantum and classical devices, 
mostly likely quantum computers enhancing more traditional classical supercomputers and High-Performance Computing (HPC) \cite{Pascuzzi:2024ktm}.
In a similar spirit, one may try to incorporate ideas from classical algorithms into the realm of quantum algorithms.

\paragraph{V. Synergy with Effective Field Theory}
In {\bf Item I} listed above, we listed the necessity of simplifying the problem. While this can mean invoking some approximation schemes, 
it is often difficult to control corrections originating from various sources.
One systematic physics framework for addressing this problem is the idea of Effective Field Theory (EFT), which implements the idea of separation of energy scales in a precise manner. Since the obstacles to scalability often originate from the existence of different energy scales, the EFT framework could be the ultimate physics reason why the problem can be tractable.\footnote{The usefulness of EFT in quantum simulations was emphasized also in Ref.~\cite{Bauer:2021gup}.} In fact, 
even in classical computations we are often 
invoking (sometimes even unknowingly) the EFT framework. EFT framework could also be a useful guideline as we scale quantum devices into those with ever-increasing number of qubits.

\qed

One should quickly emphasize that
the list above is not meant to be a complete list and the requirements on the list should not be imposed mechanically; some of the items above are not strictly necessary for finding interesting physics, and the list above should rather be thought of as ``a wishlist''
of a physicist---a wish is sometimes guaranteed, but not always.

This comment applies, for example, to {\bf Item III}.
There can be many problems in HEP that are intrinsically classical but nevertheless could be solved potentially efficiently by quantum computers. Quantum computers could be of use in solving partial differential equations arising from classical physics, or in quantum machine learning (QML) for analyzing classical data. 

Let us also point out that the items on the list are not completely unrelated. For example, {\bf Item I} on simplified models is the flip side of the {\bf Item II} on scalability, and is also related to the 
EFT framework on {\bf Item V}.
Readers are encouraged to add or delete items to the list following their own preferences.
Despite these caveats, it is my hope that this list will be of use in future exploration of 
quantum algorithms/computers in the HEP.

To make the discussion concrete, 
in the rest of this paper 
I will choose quantum Parton Shower (PS) as 
one of the problems where many of the items listed above are satisfied, at least partially.

\section{Classical Parton Showers}

Let us briefly summarize classical parton showers.
Following the spirit of the {\bf Item I}, let me explain the very basics of the parton showers in an (arguably) simplest possible setup, following the author's paper with S.~Chigusa \cite{Chigusa:2022act}. 

Let us consider the Lagrangian 
\begin{align}
\label{eq:L}
  \mathcal{L} =
  \bar{\chi} (i\slashed{\partial} - m_{\chi}) \chi
  +  ig \bar{\chi} \slashed{A}' \chi
  - \frac{1}{4} F'_{\mu\nu} F'^{\mu\nu} - \frac{1}{2} m_{A'}^2 A'_\mu A'^\mu\;,
\end{align}
describing a fermion $\chi$ coupled with a massive $U(1)$ gauge boson $A'$. While Ref.~\cite{Chigusa:2022act} considered 
a dark-sector model where $\chi$
is a dark-sector fermion and $A'_{\mu}$ is a dark-sector gauge boson\footnote{Such an extension of the dark sector is motivated by e.g.\ in the context of self-interacting dark matter (SIMP), see references in Ref.~\cite{Chigusa:2022act} for further details.}, such a specification does not matter for our discussion here, and if one prefers one can (by adding appropriate indices)
regard the gauge bosons as the $SU(3)$ gluon and $\chi$ as the quark of the standard model.

\begin{figure}[htbp]
\centering
\scalebox{1.1}{
\begin{tikzpicture}
\begin{feynman}
\vertex (a) at (0,0) {\(\chi\)};
\vertex (b) at (1, -0.5);
\vertex (f1) at (2, 0) {\(A'\)};
\vertex (c) at (1.5, -0.75);
\vertex (f2) at (2.5,-0.25) {\(A'\)};
\vertex (f3) at (4,-1) {\(A'\)};
\vertex (d) at (3, -1.5);
\vertex (g) at (4, -2) {\(\chi\)};
\diagram* {
(a) -- [fermion] (b) -- [fermion] (c) -- [fermion] (d),
(b) -- [boson] (f1),
(c) -- [boson] (f2),
(d) -- [boson] (f3),
(d) -- [fermion] (g),
};
\end{feynman}
\end{tikzpicture}
}
\caption{The simplified version of the parton shower describes a sequence of emissions of gauge bosons $A'$ from the fermion $\chi$. We can model this by splitting the time evolution into smaller steps, where a single emission either happens or not in each time step. Such a process is simulated by classical Monte Carlo techniques.}
\label{fig:emission}
\end{figure}
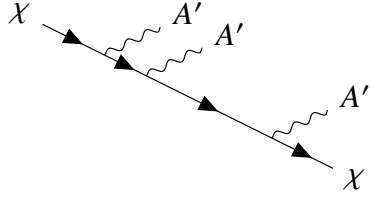

When a fermion propagates, it occasionally emits a
gauge boson $A'$, as in \cref{fig:emission}. 
We can model this process by discretizing the time steps, and if the mesh is sufficiently fine it is enough to work with an approximation where at most one emission happens during each interval. 
This problem can be treated in the 
classical Monte Carlo techniques.

The emission probability depends on the kinematics of the particles. While the proper treatment of kinematics requires an introduction of the four-momenta of particles, let us here simplify the presentation by stating that 
the relevant parameter is the energy fraction $0\le z\le 1$,
where the fermion (resp.\ gauge boson) carries $1-z$ (resp.\ $z$) of the original momentum.
The probability is determined by a distribution $f(z)$. The classical parton shower samples 
efficiently from a uniform distribution defined by $f(z) dz$,
where the function $f(z)$ can be evaluated
e.g.\ in the leading-log approximation.
In the actual sampling, one changes the variable as 
$f(z) dz = dR$, where $R$ obeys the uniform distribution in $[0, 1]$. In other words, we sample $z$ as 
$z = \mathcal{F}^{-1} (R)$,
where $\mathcal{F}(z) = \int^z dz' f(z')$
is the integral of $f$.
This sampling is not an easy problem in general, since it is difficult to obtain a closed expression for $\mathcal{F}$ and we need to evaluate its inverse based on a numerical evaluation of $\mathcal{F}$.
The classical parton shower algorithm in essence 
deal with this sampling problem
by Monte Carlo techniques.

One should be aware that many simplifications are 
involved in the explanation above.
For example, depending on the relative size between 
the twice the fermion mass $2 m_{\chi}$ and the gauge boson mass $m_{A'}$ the gauge boson may further decay into a pair of fermions, as $A'\to \chi\bar{\chi}$,
and this complicates the process by increasing the number of fermions. The basic idea of Monte Carlo sampling, however, stays the same.

\section{Quantum Interference in Parton Showers}

While classical parton shower algorithms
have been very successful, it has not properly incorporated important quantum interference effects.

To illustrate this, let us consider the same Lagrangian \eqref{eq:L} but now with $N_f$ flavors of the fermions,
\begin{align}
  \mathcal{L}_{\mathrm{dark}} =
  \sum_i \bar{\chi}_i (i\slashed{\partial} - m_{\chi_i}) \chi_i
  + \sum_{i,j} ig_{ij} \bar{\chi}_i \slashed{A}' \chi_j
  - \frac{1}{4} F'_{\mu\nu} F'^{\mu\nu} - \frac{1}{2} m_{A'}^2 A'_\mu A'^\mu\;,
\end{align}
where $i,j=1,\dots N_f$ are the flavor indices.
Note that we have chosen the basis of $\chi_i$'s so that their mass matrices are diagonal, resulting in the off-diagonal interactions $g_{i\ne j}$ in gauge interactions. In other words, the flavor of the fermion can change when the fermion emits the gauge boson: $\chi_i\to \chi_j A'$ with $i\ne j$.

\begin{figure}[htbp]
\centering
\scalebox{1.1}{
\begin{tikzpicture}
\draw (-1,1)--(-1, -2);
\begin{feynman}
\vertex (a) at (0,0) {\(\chi_i\)};
\vertex (b) at (1, -0.5);
\vertex (f1) at (2, 0.5) {\(A'\)};
\vertex (c) at (2, -1);
\vertex (f2) at (3,0) {\(A'\)};
\vertex (f3) at (3.4, -1.7) {\(\chi_j\)};
\diagram* {
(a) -- [fermion] (b) -- [boson] (f1),
(b) -- [fermion, edge label'=\(\chi_k\)] (c),
(c) -- [boson] (f2),
(c) -- [fermion] (f3),
};
\end{feynman}
\node at (4, -0.5) {$+$};
\begin{feynman}
\vertex (a) at (5,0) {\(\chi_i\)};
\vertex (b) at (6, -0.5);
\vertex (f1) at (7, 0.5) {\(A'\)};
\vertex (c) at (7, -1);
\vertex (f2) at (8,0) {\(A'\)};
\vertex (f3) at (8.4, -1.7) {\(\chi_j\)};
\diagram* {
(a) -- [fermion] (b) -- [boson] (f1),
(b) -- [fermion, edge label'=\(\chi_l\)] (c),
(c) -- [boson] (f2),
(c) -- [fermion] (f3),
};
\end{feynman}
\draw (9,1)--(9, -2);
\node at (9.25, 1) {$^2$};
\end{tikzpicture}
}
\caption{The quantum interference effects of different flavors $k\ne l$ of intermediate states,
in the process $\chi_i \to \chi_j A' A'$.}
\label{fig:interference}
\end{figure}
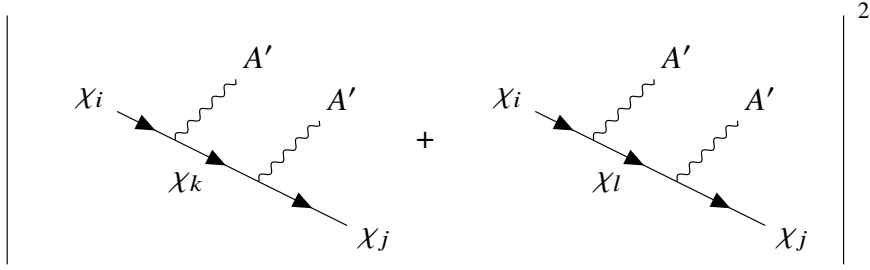

Suppose now that we have a sequential emitting process
$\chi_i\to \chi_k A'$ and $\chi_k \to \chi_j A'$,
so that we have $\chi_i\to \chi_j A' A'$.
In this process, we need to take care of the quantum interference effects for the different flavors of intermediate states, as shown in \cref{fig:interference}.
This is the quantum interference effect in the quantum parton shower.

In view of the enormity of the parton-shower literature, it is surprising that this quantum interference has not been fully incorporated (see e.g.\ \cite{Dokshitzer:1978hw})---the effect has partly been incorporated but always involves some approximation schemes, e.g.\ the large $N_c$ limit where the Feynman diagrams become planar.

We emphasize that the 
quantum interference effect has not been fully incorporated until quite recently, even on classical computers. This is simply because the Monte Carlo methods are classical algorithms and hence by definition do not incorporate quantum effects. 
Quantum algorithms for quantum parton showers is a new subject which is currently actively 
studied, see e.g.\ 
Refs.~\cite{Bepari:2020xqi,Li:2021kcs,Bepari:2021kwv,Macaluso:2021ngq,Gustafson:2022dsq} for a sample of references.

While one may expect such quantum interference effects are suppressed by the entries of the Cabibbo–Kobayashi–Maskawa (CKM) matrix and to be not too large for
standard-model processes at LHC, the analysis of dark sector models in 
Ref.~\cite{Chigusa:2022act} demonstrates that such effects can be dramatic
and can actually be observed in future colliders, at least in principle.
In this sense, we have satisfied the {\bf Item III}.

Let us also point out that our discussion here
fits well with the EFT philosophy of splitting the problem according to their energy scales, as in {\bf Item V}. 
In collider simulations, the parton shower is 
the intermediate stage between the Matrix Elements (ME) stage and hadronization stage (for QCD in SM).
At high energies (at $\sim O(100)\rm{GeV}$--$O(1){\rm TeV}$), one starts with the high energy where new known physics may play important roles.
This region, however, is the perturbative region of the QCD where
perturbative techniques are available.
Down to the scale of $\sim O(100){\rm MeV}$--$O(1){\rm GeV}$, the quarks hadronize, which is a 
highly non-perturbative process but can be well-modeled partly thanks to experimental inputs.
The factorization theorems \cite{Collins:1987pm,Sterman:1995fz,Jaffe:1996zw,Bauer:2002nz} guarantee that we can in principle 
separate parton showers from the ME and hadronization, thereby making it possible to restrict the size of the problem.

The first quantum parton shower algorithm was 
proposed in \cite{Bauer:2019qxa}.
The quantum circuit was subsequently simplified by
incorporating mid-circuit measurements and dynamic circuits
\cite{Deliyannis:2022uyh}.

To make the algorithm more realistic, one needs to 
incorporate the kinematical data which was absent in \cite{Bauer:2019qxa,Deliyannis:2022uyh}.
This is where the quantum-classical hybrid naturally comes in, since
it is much more efficient to hold the kinematics information (the size of the four-momenta)
in the classical CPUs---the quantum circuits could concentrate on keeping track of the presence/absence of the emission processes,
where quantum interference effects play crucial roles.
While it is non-trivial to ensure that such a splitting into classical and quantum parts works consistently,
such a hybrid algorithm was implemented in my paper with
C.W.~Bauer and S.~Chigusa \cite{Bauer:2023ujy}.

\begin{figure}[htbp]
\centering
\scalebox{1.1}{
\begin{tikzpicture}
\draw [->] (0.2,-0.3) -- (1, -0.7);
\draw [->] (1.2,-0.8) -- (2, -1.2);
\draw [->] (2.2,-1.3) -- (3, -1.7);
\draw (0.5,-0.65) node {$1$};
\draw (1.3,-1.15) node {$1-z$};
\draw (2.0,-1.75) node {$1-z-z'$};
\draw [->] (1,-0.3) -- (1.7, 0.4);
\draw [->] (2,-0.8) -- (2.7, -0.1);
\draw (1.3,0.2) node {$z$};
\draw (2.3,-0.3) node {$z'$};
\begin{feynman}
\vertex (a) at (0,0) {\(\chi\)};
\vertex (b) at (1, -0.5);
\vertex (f1) at (2, 0.5) {\(A'\)};
\vertex (c) at (2, -1);
\vertex (f2) at (3,0) {\(A'\)};
\vertex (f3) at (3.4, -1.7) {\(\chi\)};
\diagram* {
(a) -- [fermion] (b) -- [boson] (f1),
(b) -- [fermion] (c),
(c) -- [boson] (f2),
(c) -- [fermion] (f3),
};
\end{feynman}
\node at (4, -0.5) {or };
\draw [->] (5.2,-0.3) -- (7, -1.2);
\draw [->] (7.2,-1.3) -- (8, -1.7);
\draw (5.7,-0.75) node {$1$};
\draw (7.0,-1.75) node {$1-z'$};
\draw [->] (7,-0.8) -- (7.7, -0.1);
\draw (7.3,-0.3) node {$z'$};
\begin{feynman}
\vertex (A) at (5,0) {\(\chi\)};
\vertex (B) at (6, -0.5);
\vertex (C) at (7, -1);
\vertex (F2) at (8,0) {\(A'\)};
\vertex (F3) at (8.4, -1.7) {\(\chi\)};
\diagram* {
(A) -- [fermion] (C),
(C) -- [boson] (F2),
(C) -- [fermion] (F3),
};
\end{feynman}
\end{tikzpicture}
}
\caption{Whether or not an emission happens at a particular Monte Carlo step affects the 
kinematics, and hence the emission probabilities through RG running, in all the subsequent steps.
Such a history dependence can be incorporated into the 
quantum circuit either by preparing a history counter or invoking a dynamic circuit.
}
\label{fig:hysterisis}
\end{figure}
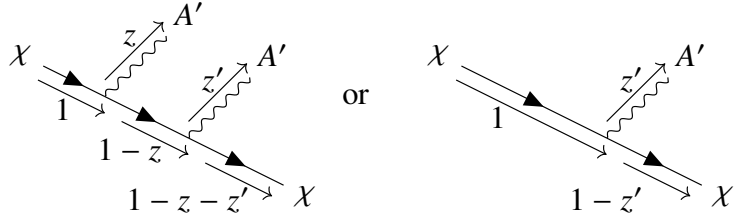

One of the subtleties of kinematical parton showers is that the result at a particular step of the PS affects all the subsequent steps. For example, in \cref{fig:hysterisis}
one finds that the emission/non-emission at one step affects the kinematics of the subsequent steps. Moreover, the emission probability depends on the gauge coupling constant, which in turn depends on the energy scale due to the renormalization group (RG) running. This means that the emission probability in itself depends on the history of all the previous steps. The quantum circuit will then necessarily be a
dynamic circuit, where the results of the measurement of some qubits will be feed-forwarded
to affect the quantum circuits in later steps (e.g.\ by changing rotation angles in controlled-rotation gates). 

Along this line of research we realized that it is very useful to 
incorporate the veto algorithm well-known in the literature.
While no details are needed for the purposes of this paper,
let us quickly summarize the main points of the veto algorithm.

Suppose that we have a probability distribution $f(\bm{q}, \bm{c}; z)$, which depends on 
both quantum state $|\bm{q}\rangle$ as well as classical data $\bm{c}$.
When we sum over the quantum states, we have a classical distribution $f(\bm{c}; z)= \sum_{\bm{q}} \alpha_{\bm{q}}^2 f(\bm{q}, \bm{c}; z)$, where we have assumed a density matrix $\rho = \alpha_{\bm {q}} |\bm{q}\rangle \langle \bm{q} |$.
When we apply the naive Monte Carlo techniques, we end up numerically evaluating the \emph{inverse} of the function
\begin{align}
    \mathcal{F}(\bm{c}; z)
= \int^z dz' f(\bm{c}; z')
= \sum_{\bm{q}} \alpha_{\bm{q}}^2 \int^z dz' f(\bm {q}, \bm{c}; z') \;,
\end{align}
which in itself is a linear combination of $2^{n}$ functions, where $n$ is the size of the qubit $|\bm{q}\rangle$; this becomes intractable as $n$ becomes large.
In the veto algorithm, one replaces 
the function $f(\bm{q}, \bm{c}; z')$
by a simpler function $f^{\rm over}(\bm{c}; z')$ which satisfies that
(1) $f^{\rm over}$ is independent of the quantum state $\bm{q}$ 
(2) $f^{\rm over}(\bm{c}; z')\ge f(\bm{q}, \bm{c}; z')\ge 0$ for all $\bm{q}, \bm{c}$ (3) the integral $\mathcal{F}^{\rm over}(\bm{c}; z')
=\int^z dz' f^{\rm over}(\bm{c}; z')$
as well as its inverse can be evaluated analytically. The Monte Carlo procedure
for $f^{\rm over}$ can then be easily performed since we no longer have quantum superposition and the inverse of $F^{\rm over}$ is analytically known. Of course, one needs to correct for the difference between $f^{\rm over}$ and $f$, but that can be taken into account by a veto procedure, which in particular does not require evaluation of the inverse function $\mathcal{F}^{-1}$. While care is needed in 
implementing the veto procedure for quantum circuits, this was 
successfully implemented in the limit $f(\bm{q}, \bm{c}; z) \ll 1$, see Ref.~\cite{Bauer:2023ujy} for details.

After we incorporated the veto algorithms, the quantum circuit simplifies dramatically---when we have $N$ PS steps and $N_f$
flavors then the number of qubits 
scales as $O( (\log_2 N_f) N)$
and the number of two-qubit gates as $O(N_f^2 N)$, both scaling as linear in the PS steps $N$.
In fact, if we assume that the input state to the parton shower is not highly entangled then we can efficiently simulate the parton showers by classical computers for some reasonable values of $N_f$ and $N$.
While our discussion still is in simplified models and there is room for further exploration, 
it seems fair to expect that the problem is fully scalable into the future, thus satisfying the requirements of {\bf Item II}.

\begin{table}[htbp]
\caption{Checklist for the quantum parton showers, with doubled circle (circle)
for ``very good'' (``good'').
The grades here reflect my biases, and I am admittedly somewhat generous in my estimate here; as I explained in the main text there is much room for further study.}
\centering
\begin{tabular}{|p{4cm}|c|p{9cm}|}
\hline
   I. Availability of Simplified Models 
   & $\doublecirc$ 
   & We can start with the simplified models \cite{Bauer:2019qxa,Chigusa:2022act},
    which already incorporate the important quantum interference effects. 
   \\
    \hline
   II. Scalability
   & $\doublecirc$/? 
   & In some reasonably realistic scenarios, the number of both qubits and gates  grows as 
    $O(N)$ for $N$ PS steps. One therefore expects that the problem is scalable.
    One needs a better understanding of the relevant physics models (in SM and beyond), however, to better estimate e.g.\ the precisions required for extracting important physics. 
   \\
\hline
   III. Importance of Quantum Effects
   & $\doublecirc$/$\bigcirc$ 
   & The quantum interference effect has been missed in classical PS algorithms, and properly incorporating itself in PS algorithms was in itself an achievement. Such quantum effects could play significant roles as pointed out in Ref.~\cite{Chigusa:2022act}. Further study is necessary, however, whether or not such effects play important roles
   in the discussions of SM at LHC or future colliders,
   when we assume no new physics beyond the SM.
   \\
   \hline
   IV. Interface to Classical Computations
   & $\doublecirc /?$ 
   & A quantum-classical hybrid algorithm 
    based on dynamic circuits has successfully been implemented in QPS. 
    It remains to be seen, however, how the quantum parton shower algorithms discussed here can be combined with 
    state-of-the-art classical programs.
   \\
\hline
   V. Synergy with Effective Field Theory
   & $\doublecirc$ 
   & The separation of Parton Shower from 
    ME/hadronization computations at 
    other energy scales is partly guaranteed by the factorization theorem.
   \\
\hline
\end{tabular}
\end{table}

\section{Lessons for Further Exploration}

In this paper, we first discussed a wishlist when
we consider applications of quantum computers/algorithms to 
HEP (and more general problems in physics). We then turned to 
quantum parton showers as an example of a problem
where such criteria are at least partially satisfied.

The discussion of this paper raises several important questions
for quantum PS, for example:

\begin{itemize}
\item Even apart from quantum simulations, it is an important physics question to better understand the 
importance of quantum interference effects, and
to give an estimate of such effects. 

\item We have not fully addressed the question of
the interface with classical computers. While this is partly incorporated in our quantum-classical hybrid algorithm, we eventually hope to make contact with state-of-the-art quantum PS shower simulators, such as 
PYTHIA \cite{Bierlich:2022pfr}, Herwig \cite{Bahr:2008pv}, Sherpa \cite{Sherpa:2024mfk}.

\end{itemize}

In addition to the interest in the physics of the quantum PSs themselves, one would hope to regard the PS as a test case for applying quantum algorithms/computers to many more interesting problems in HEP. In this spirit, let us make some general remarks, which may be regarded as the lessons from the developments in the quantum PS:

\begin{itemize}
   \item There are some problems in HEP
   where quantum interference effects have not been fully incorporated.

   \item In addition to providing powerful computational resources,
   quantum computers stimulate the development of physics problems themselves 
   (as we have mentioned in the first point above on quantum PSs).
   In this sense, the interaction between 
   HEP and Quantum can work in both directions.

   \item While quantum-classical algorithms have already been used intensively in the literature (e.g.\ in Variational Quantum Eigensolver (VQE) and Quantum Machine Learning), discussion of QPS suggests that 
   more opportunities are present once we study 
   dynamic circuits, where the results of the measurements are feed-forwarded to the rest of the quantum circuits (see e.g. \ \cite{Baumer:2023vrf} for an example of a recent discussion of dynamic circuits).  
   
   \item Quantum algorithms can benefit tremendously from existing classical algorithms, as we have seen in the example of the veto algorithm. The ``translation'' of classical algorithms to quantum algorithms, however, is often not trivial, due to the existence of quantum entanglement in the quantum circuits.
 
   \item Dedicated community efforts will be required for upgrading existing classical algorithms fully into the quantum or quantum-classical-hybrid regimes.\footnote{I would imagine that the present-day parton-shower (and event-generator) programs such as PYTHIA, Herwig, Sherpa, will one day be replaced by their quantum counterparts, which may naturally be called QPYTHIA, QHerwig, QSherpa.}
\end{itemize}

\section{Acknowledgements}

The author would like to C.W.~Bauer and S.~Chigusa
for collaboration in \cite{Chigusa:2022act,Bauer:2023ujy}, on which this talk is based. He would also like to thank Y.~Iiyama and V.~Pascuzzi for related discussions.
The content of this paper was presented at various conferences and he would like to thank the audience for feedback. He would like to thank the organizers of the QCHSC24 conference for providing a stimulating and tropical environment.
This research was supported in part by the JSPS Grant-in-Aid for Scientific Research (No.\ 20H05860, 23K17689, 23K25865),  by JST, Japan (PRESTO Grant No.\ JPMJPR225A, Moonshot R\&D Grant No.\ JPMJMS2061), and by IBM-UTokyo lab.

\bibliographystyle{JHEP}
\bibliography{bib}

\end{document}